\documentclass[aps,prb,twocolumn,showpacs,preprintnumbers,amsmath,amssymb,supers
criptaddress]{revtex4}
\usepackage{dcolumn} \usepackage{bm} \usepackage{graphicx}
\usepackage{graphics} \usepackage{epsfig} \usepackage{float}
\usepackage{amsmath} \usepackage{amssymb} \usepackage{bm}

\begin{document}


\title{High-Temperature Superconductivity in Eu$_{0.5}$K$_{0.5}$Fe$_{2}$As$_{2}$  }
\author{H. S. Jeevan}
\email{Jeevan.Hirale@phys.uni-goettingen.de} \affiliation{I. Physik. Institut,
Georg-August-Universit\"{a}t G\"{o}ttingen, D-37077 G\"{o}ttingen,
Germany}

\author{Z. Hossain}
\affiliation{Department of Physics, Indian Institute of Technology,
Kanpur 208016, India}
\author{Deepa Kasinathan}
\affiliation{Max-Planck Institute for Chemical Physics of Solids,
D-01187 Dresden, Germany}
\author{H. Rosner}
\affiliation{Max-Planck Institute for Chemical Physics of Solids,
D-01187 Dresden, Germany}
\author{C. Geibel}
\affiliation{Max-Planck Institute for Chemical Physics of Solids,
D-01187 Dresden, Germany}
\author{P. Gegenwart}
\affiliation{I. Physik. Institut, Georg-August-Universit\"{a}t
G\"{o}ttingen, D-37077 G\"{o}ttingen, Germany}

\date{\today}
       
\begin{abstract}
Subsequent to our recent report of SDW type transition at 190~K and
antiferromagnetic order below 20~K in EuFe$_{2}$As$_{2}$, we have 
studied the effect of
K-doping on the SDW transition at high temperature and AF order at
low temperature. 50\% K doping suppresses the SDW transition and
in turn gives rise to high-temperature superconductivity below 
T$_{c}$ = 32~K,
as observed in the electrical resistivity, AC susceptibility as well
as magnetization. A well defined anomaly in the specific heat
provides additional evidence for bulk superconductivity. 
\end{abstract}


\maketitle

\section{\label{sec:level2}Introduction}

Intense research for the past few months have lead to the discovery of
several new superconductors with $T_c$ as high as 52~K in the iron
arsenide compounds
\cite{Takahashi,Chen,GFChen,Rotter,Ding,Wu,TegelSC,Li}. Most
intriguing among these superconductors is the fact that
superconductivity occurs close to the magnetic-nonmagnetic
borderline probably indicating an unconventional (non-phononic)
pairing mechanism. While the debate about the mechanism and the
conventional versus unconventional nature continues, we try to
explore more materials with the aim to assist in sorting out this
debate through more experiments on related compounds. Recently, we
have shown that EuFe$_{2}$As$_2$ shows a spin-density-wave (SDW)
type transition at 190~K \cite{Jeevan} similar to that of
SrFe$_{2}$As$_2$ \cite{Krellner}. One may thus expect that
appropriate doping or application of pressure suppresses the SDW,
leading to superconductivity at the magnetic-nonmagnetic borderline.
Taking clue from the results of K-doped SrFe$_{2}$As$_2$ \cite{Chen}
we may expect that 50\% K doped EuFe$_{2}$As$_2$ shows
superconductivity. We have prepared 50\% K doped EuFe$_{2}$As$_2$
and measured its low-temperature electrical resistivity, magnetic
susceptibility and specific heat. We indeed find compelling
experimental evidence for bulk superconductivity in K-doped
EuFe$_{2}$As$_2$. Details of the experimental procedures and results
are discussed below.

\section{Methods}
\subsection{\label{sec:level2}Experimental}

Polycrystalline samples of Eu$_{0.5}$K$_{0.5}$Fe$_{2}$As$_{2}$ were
synthesized by solid state reaction method. Stoichiometric amounts
of the starting elements Eu (99.99 $\%$), K (99.9$\%$), Fe
(99.9$\%$) and As (99.999$\%$) were taken in an Al$_{2}$O$_{3}$
crucible which was then sealed in a Ta-crucible under Argon
atmosphere. The sealed Ta-crucibles were slowly heated to
$900^\circ$C and sintered for 24 hours. After first heat treatment,
the samples were thoroughly grounded and pressed into pellets and
subjected to a second heat treatment. The sample preparation and
handling was carried out inside a glove box (O $<$ 1 ppm , H$_2$O
$<$ 1 ppm). Electrical resistivity and specific heat were measured
using a Physical Properties Measurement System (PPMS, Quantum
Design, USA). The AC-Susceptibility was measured using home-made
system while for the dc magnetic susceptibility a Quantum Design
MPMS was used.

\subsection{Theory}
In order to understand the influence of the K doping in
EuFe$_{2}$As$_{2}$, we have performed density functional band
structure calculations using a full potential code FPLO\cite{fplo}
within the local (spin) density approximation (L(S)DA).  Additionally,
we have included the strong Coulomb repulsion in the Eu 4$f$ orbitals
on a mean field level using the LSDA+$U$ approximation applying the
atomic-limit double counting scheme\cite{czyzyk}.  We applied the
Perdew-Wang\cite{perdew} flavor of the exchange correlation potential,
self-consistency was reached on a well converged $k$- mesh.  The
structural parameters and the value of $U$ = 8 eV for the Eu 4$f$
states were kept identical to our previous work\cite{Jeevan} on the
undoped system throughout all calculations.

\section{\label{sec:level2}Results and Discussion}
\subsection{Experiment}

The single-phase nature of the sample is confirmed using powder
x-ray diffraction. Impurity phases amount to less than 5 $\%$. The
sample crystallizes in the ThCr$_2$Si$_2$ type tetragonal structure
with lattice parameters $a = 3.8671(3){\AA}$ and $c = 13.091(3){\AA}$. The
lattice parameter values as expected, are in between those 
of EuFe$_{2}$As$_2$ and KFe$_{2}$As$_2$. EDAX analysis
reveals, that the composition of the sample is close
to the expected 0.5: 0.5: 2: 2.

\begin{figure}
\includegraphics[width=8.5cm]{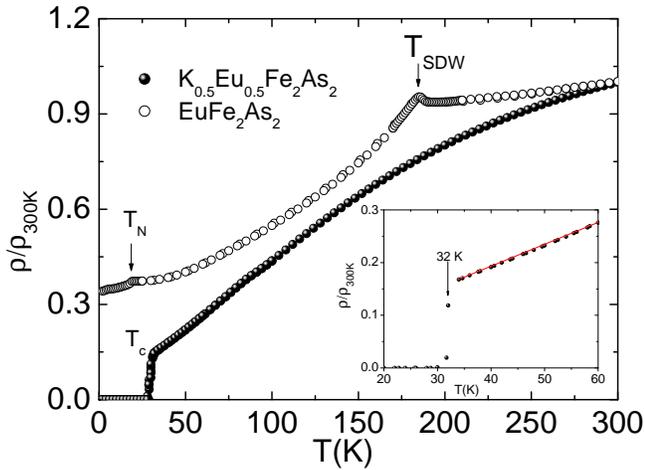}
\caption{\label{fig 1}  Temperature dependence of the electrical
resistivity of polycrystalline EuFe$_{2}$As$_2$ and
Eu$_{0.5}$K$_{0.5}$Fe$_{2}$As$_{2}$. The arrows indicate anomalies
at T$_{SDW}=190$~K and T$_{c}=30$~K. The lower inset shows the
electrical resistivity data in temperature range between 20 and
60~K, the red line indicates the linear decrease between 60 and 32
K.}
\end{figure}

Fig.~1 shows the temperature dependence of the electrical
resistivity for EuFe$_{2}$As$_2$  and
Eu$_{0.5}$K$_{0.5}$Fe$_{2}$As$_{2}$ samples. Undoped
EuFe$_{2}$As$_2$ exhibits a clear drop near 190~K due to SDW
instability \cite{Jeevan,Raffius,ren} and structural phase transitions
from tetragonal to orthorhombic symmetry \cite{Tegel}. In the
K-doped sample Eu$_{0.5}$K$_{0.5}$Fe$_{2}$As$_{2}$, the
high-temperature anomaly due to the SDW formation is completely
suppressed. By contrast, a superconducting transition is observed at
$T_c\approx 32$~K. The lower inset of Fig. 1 shows the low
temperature part of resistivity. Note, that a linear temperature
dependence is found above the superconducting transition.

\begin{figure}
\includegraphics[width=8.5cm]{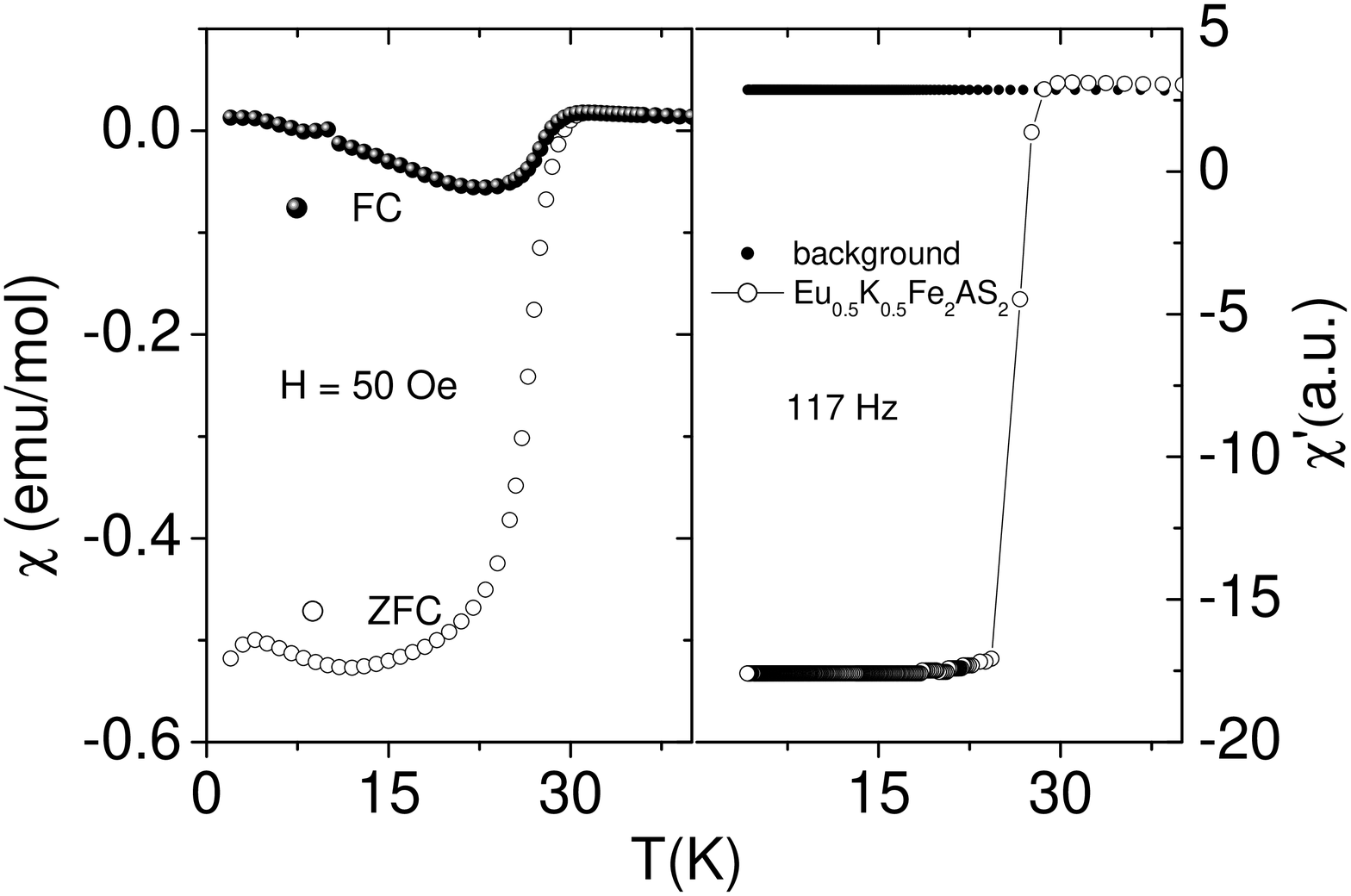}
\caption{\label{fig 2} Temperature dependence of (a) DC magnetic
susceptibility for zero-field-cooled (ZFC) and field-cooled (FC)
experiments in an applied field of~50 Oe and (b) the real part of
the AC susceptibility measured with oscillating frequency of 117 Hz.
The dots indicate the background signal.}
\end{figure}

Further confirmation of the superconducting transition at 32~K comes
from AC magnetic susceptibility (Fig. 2) which shows a clear
diamagnetic signal. DC magnetic susceptibility measured under 
zero-field-cooled (ZFC) and field-cooled (FC) conditions also confirm
superconductivity. The ZFC diamagnetic signal below the
superconducting transition is of the size expected for perfect
diamagnetism of Pb. FC measurements also exhibit diamagnetism, the
diamagnetic signal being 10\% of the ZFC signal. The hysteresis
between the ZFC and FC susceptibility in the superconducting state
is characteristic for type-II superconductors. We also observe a
signature for magnetic ordering of the Eu-moments in the magnetic
susceptibility below about 10~K.

\begin{figure}
\includegraphics[width=8.5cm]{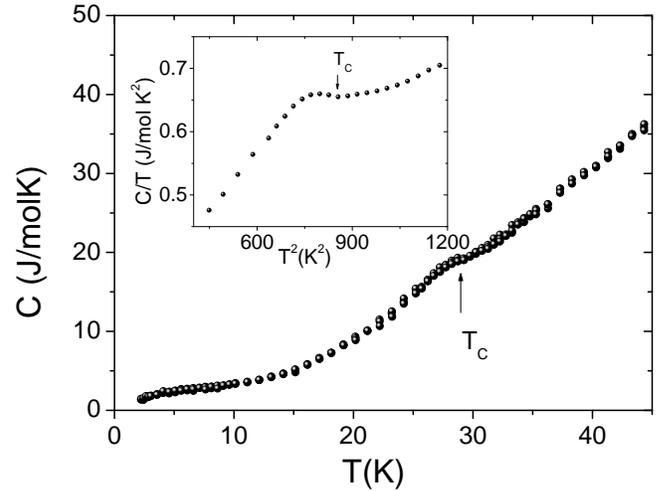}
\caption{\label{fig 3}  Temperature dependence of the specific heat
of Eu$_{0.5}$K$_{0.5}$Fe$_{2}$As$_{2}$ in the temperature range
4~K$\leq T\leq 40$~K. The inset displays $C/T$ vs $T^2$.}
\end{figure}

Since even 10 - 20\% superconducting volume fraction could provide
zero resistivity as well as a diamagnetic signal similar to that of
a pure superconductor, low-temperature specific heat measurements
were carried out to establish the bulk nature of superconductivity,
as well as to probe the magnetism of the Eu$^{2+}$ ions. The main
part of Fig.~3 shows the temperature dependence of the specific heat
plotted as $C$ vs $T$ in the temperature range 2 - 40~K.
We observe a small but well defined anomaly associated with the superconducting
transition providing clear evidence for the bulk nature of
superconductivity. It is even more pronounced in $C/T$ vs $T^2$
shown in the inset. From  this plot, one can estimate a step $\Delta C/T_{c}
\approx $ 50 mJ/mol K$^{2}$ in the specific heat at T$_{c}$. Unfortunately
because of the presence of the magnetic contribution
from Eu-moments, it is difficult to correctly estimate the true
normal-state $\gamma$ value. Since $\gamma$ of SrFe$_{2}$As$_2$ is
reported as 10~mJ/mol K$^{2}$ we believe the true gamma value of
EuFe$_{2}$As$_2$ will be of the same order of magnitude, considering
the fact that Eu is divalent in the absence of charge fluctuations
and/or Kondo effect. 
Compared to this $\gamma$-value, the size of  $\Delta C/T_{c}$ we found
in Eu$_{0.5}$K$_{0.5}$Fe$_{2}$As$_{2}$ is quite large, suggesting a strong 
coupling scenario. 
 Even though the mechanism
for the superconductivity is not yet settled, the well defined
specific heat anomaly of this magnitude provides confidence about
the bulk nature of superconductivity in
Eu$_{0.5}$K$_{0.5}$Fe$_{2}$As$_{2}$. This is so far the best
evidence of the bulk nature of superconductivity in FeAs-family of
superconductors. Such anomaly was absent in LaO$_{1-x}$F$_{x}$FeAs
compounds \cite{Mu,Athena} and only a weak anomaly was
observed in superconducting SmO$_{1-x}$F$_{x}$FeAs \cite{Ding}. At
low temperatures only a very broad bump suggesting a suppressed and
broadened kind of magnetic order of Eu-moments very likely only
a short range one, was observed below 10
K in contrast to a very pronounced lambda-type anomaly at 20 K in
EuFe$_{2}$As$_2$ \cite{Jeevan}.

\subsection{Theory}

To study the changes in the electronic stucture of EuFe$_{2}$As$_{2}$
due to a 50\% doping of K on the Eu site, we constructed supercells of
the undoped system and replaced half of the Eu ions with K ions.  To
account for the strong Coulomb repulsion within the Eu 4$f$ orbitals a
typical $U$ value (according to XAS and photoemmission experiments) of
$U$ = 8\,eV has been chosen for the LSDA+$U$ calculations. A variation
of $U$ within the physically reasonable range of 6...10\,eV does not
change our conclusions since this variation has negligible influence
on the states at the Fermi level that are relevant for the
superconductivity.

\begin{figure}[t]
\begin{center}
\includegraphics[%
  clip,
  width=7cm,
  angle=-90]{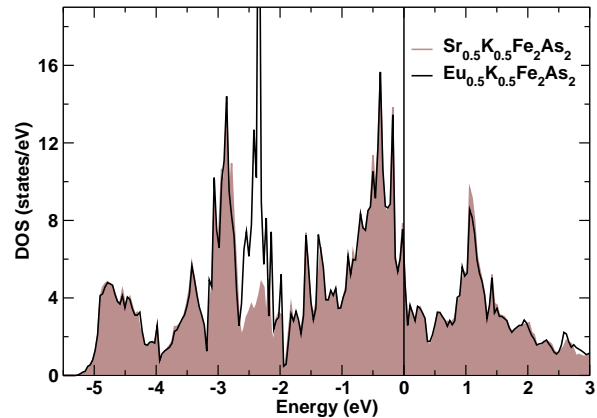}
\end{center}
\caption{\label{dos}(Color online) Comparison of the total DOS for
  Eu$_{0.5}$K$_{0.5}$Fe$_{2}$As$_{2}$ (LSDA+$U$, with non-magnetic Fe
  sites) and Sr$_{0.5}$K$_{0.5}$Fe$_{2}$As$_{2}$ (LDA). The spin-up
  and spin-down DOS with non magnetic Fe sites for the
  Eu$_{0.5}$K$_{0.5}$Fe$_{2}$As$_{2}$ has been added together to make
  the comparison with the Sr analogue easier.  The large peak at -2.4
  eV belongs to the spin-up Eu 4$f$ states. The unfilled spin-down Eu
  4$f$ states are pushed to around 9.5 eV above the Fermi level.  }
\end{figure}

The calculated total density of states (DOS) of
Eu$_{0.5}$K$_{0.5}$Fe$_{2}$As$_{2}$ is shown in Fig.~\ref{dos} in
comparison to that of Sr$_{0.5}$K$_{0.5}$Fe$_{2}$As$_{2}$. Except for
the localized Eu 4$f$ states at around -2.4 eV, the DOS of both these
systems are quite similar.
\footnote{In order to compute the non-magnetic electronic structure of
  Eu$_{0.5}$K$_{0.5}$Fe$_{2}$As$_{2}$ relevant for the superconducting
  ground state, we stabilized a selfconsistent solution with
  negligible polarization at the Fe sites by setting the initial spin
  split of Fe to zero. Both spin channels have been added afterwards
  for the comparison with the unpolarized calculation in the case of
  Sr$_{0.5}$K$_{0.5}$Fe$_{2}$As$_{2}$.} As in the undoped
EuFe$_{2}$As$_{2}$, the Eu ions are in a stable 2+ state with a half
filled 4$f$ shell, and the position of the localized 4$f$ level
remains basically unchanged.\cite{Jeevan}

Our previous work\cite{Jeevan} on the undoped system indicates that
the Eu and Fe sublattices are quite decoupled and the ordering of the
Fe sublattice at 190 ~K does not influence the ordering of the Eu
moments at 20 K. Thus, the ordering of the Eu spin moments is mainly
due to Eu-Eu interaction. In our Eu$_{0.5}$K$_{0.5}$Fe$_{2}$As$_{2}$
supercell, every other Eu site is replaced by a non-magnetic K$^{2+}$
ion. Although the influence of disorder can not be accessed this way,
we try to get a rough estimate for the reduction of the effective
coupling comparing calculated total energies for different Eu spin
arrangements (aligned and anti-aligned Eu spins). The difference in
energy from these calculations are then mapped to an Heisenberg model
to obtain the value of the effective exchange constant
(J$_{eff}$). The value of the J$_{eff}$ for the undoped and the 50\% K
doped EuFe$_{2}$As$_{2}$ systems are 2\,K and 0.3\,K, respectively.
Thus, replacing every other magnetic Eu$^{2+}$ ion with a non-magnetic
K$^{2+}$ ion reduces the effective Eu-Eu exchange strongly.
However, even without this strong reduction of the effective exchange, the
random filling of the square with only 50\%  magnetic Eu ions 
should be sufficient to supress long range magnetic order by itself.

\section{Conclusion}

To summarize, we have been successful in making a single phase
sample of 50\% K doped EuFe$_{2}$As$_2$. Our comprehensive
investigation of the electrical resistivity, AC and DC magnetic
susceptibility and specific heat shows the suppression
of the SDW transition and evidence for bulk type-II
superconductivity below T$_c = 32$~K. Removal of 50\% of Eu
however has substantially broadened the  Eu order 
which is then very likely only a short range one; and shifted 
the latter to below 10~K. Further experiments are planned
to probe the magnetism of Eu-ions as well as to investigate the
interplay of Eu-magnetism with superconductivity.

The authors would like to thank K. Winzer for help with the
AC-susceptibility measurements. We acknowledge financial support by
the DFG Research Unit 960 and BRNS (grant no. 2007/37/28).

\end{document}